\documentclass[12pt]{article}
\usepackage{latexsym, epsfig, graphics}
\newcommand{\be}{\begin{equation}}
\newcommand{\ee}{\end{equation}}
\newcommand{\bq}{\begin{eqnarray}}
\newcommand{\eq}{\end{eqnarray}}
\newcommand{\intsigt}{\int d\tau \int d\sigma}
\newcommand{\cue}{{\bf q}}
\newcommand{\parsig}{\partial_{\sigma}}
\newcommand{\parta}{\partial_{\tau}}
\newcommand{\bfwh}{{\bf y}}
\newcommand{\sigta}{\sigma,\tau}
\newcommand{\sigtap}{\sigma',\tau}
\newcommand{\zhee}{{\bf z}}
\begin{document}
\begin{titlepage}
\today          \hfill 
\begin{center}
%hfill    LBNL-xxix \\
%          \hfill    UCB-PTH-xx/xx \\

\vskip .5in

{\large \bf Covariant And Local Field Theory On The World Sheet}
\footnote{This work was supported by the Director, Office of Science,
Office of Basic Energy Sciences, of the U.S. Department of Energy under
Contract No. DE-AC02-05CH11231}
\vskip .50in

%alternate footnote for faculty:
%\footnote{This work was supported in part by the Director, Office of 
%Energy Research, Office of High Energy and Nuclear Physics, Division of 
%High Energy Physics of the U.S. Department of Energy under Contract 
%DE-AC03-76SF00098 and in part by the National Science Foundation under 
%grant PHY-0098840.}

\vskip .5in
Korkut Bardakci\footnote{Email: kbardakci@lbl.gov}

{\em Department of Physics\\
University of California at Berkeley\\
   and\\
 Theoretical Physics Group\\
    Lawrence Berkeley National Laboratory\\
      University of California\\
    Berkeley, California 94720}
\end{center}

\vskip .5in

\begin{abstract}
In earlier work, using the light cone picture, a world sheet field
theory that sums planar $\phi^{3}$ graphs was constructed and developed. Since
 this theory is both non-local and not explicitly Lorentz invariant,
it is desirable to have a covariant and local alternative. In this paper,
we construct such a covariant and local world sheet field theory, and
show that it is equivalent to the original non-covariant version.

\end{abstract}
\end{titlepage}%THIS PAGE (PAGE ii) CONTAINS THE LBL DISCLAIMER

\newpage
\renewcommand{\thepage}{\arabic{page}}
\setcounter{page}{1}
%THIS IS PAGE 1 (INSERT TEXT OF REPORT HERE)
\noindent{\bf 1. Introduction}
\vskip 9pt

In this article, we continue   the project
 of putting planar field theory on the world sheet.
The new feature of the present work is that for the first time,
an explicitly Lorentz invariant treatment is presented. This is an important
advance, since,
most of the earlier work, following t'Hooft's seminal paper [1],
dealt with the planar $\phi^{3}$ field theory in the light cone frame.
Although it had many convenient features, this approach  had the 
disadvantage of not being explicitly Lorentz covariant.
 A manifestly
covariant treatment was sketched in [2], but this was done in the framework
of the first quantized approach. As explained in section 2, on the world
 sheet,
a typical graph of the planar $\phi^{3}$ theory can be pictured as a
collection of parallel solid lines, which forms the boundaries of the
propagators (Fig.1).
\begin{figure}[t]
\centerline{\epsfig{file=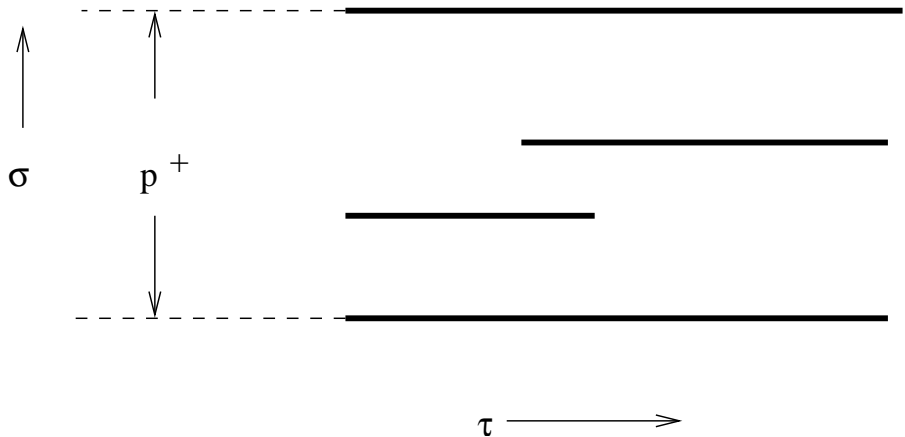, width=12cm}}
\caption{A Typical Graph}
\end{figure}
 The  momenta carried by the graph
 run through these lines, and in
the first quantized approach, these are treated as fields and quantized.
There is, however, a drawback to this approach: The interaction takes
place by the creation and annihilation of the solid lines (boundaries),
and this is difficult to accommodate in a first quantized theory. In
reference [3], this problem was overcome by developing a second quantized
theory, which naturally incorporates 
 the creation and destruction of the boundaries. One is able to
make quite a bit of progress with this  approach; for
example,
in  [4], using the mean field approximation, this model was
shown to lead to string formation.

The second quantized theory described above was constructed in the light cone
framework, and although it is intrinsically Lorentz invariant,
 it has again the disadvantage of not being manifestly
covariant.
 Any approximation scheme, such as the mean field method, is
likely to spoil Lorentz invariance. The model is also non-local on
the world sheet, which complicates matters. Also, there is a spurious
singularity at $p^{+}=0$ that is typical of the field theories in the
light cone coordinates. This singularity  drops out of the physical
quantities in the exact theory, but it is problematic in any approximation
scheme. Finally, the theory has to be renormalized in higher dimensions,
and  renormalization a non-covariant theory is notoriously
difficult\footnote{See [5] for a regulator that preserves Lorentz
 invariance.}.

The approach we will develop does not suffer from any of the difficulties
listed above. It is a second quantized theory that is
explicitly covariant and local on the world sheet. Needless to say,
 it does not
have spurious singularities. It is therefore well suited to an
approximation scheme such as the mean field method. To keep the paper to
a reasonably length, we will not attempt to carry out such a calculation
here, and leave it as an interesting project for future research.

A good portion of the paper is devoted to the review of the model in
the light cone picture. This is necessary to make the paper self
contained, but also, in reviewing the light cone picture, we will develop
the tools needed for the covariant approach. Finally, we prove the validity
of the covariant approach by showing that it is equivalent to the well
established light cone model.

In section 2, we briefly review the world sheet picture of the graphs of the
planar $\phi^{3}$ theory in the light cone variables. Section 3 is
a review of the first quantized world sheet formalism developed in [2], which
reproduces the free massless
 propagators of the light cone picture. In particular,
we emphasize the analogy with the one dimensional electrostatics. In this
analogy, the momenta correspond to the electrostatic potential, which is
generated by charges residing on the boundaries. These charges will play 
an important role in the next section, where we construct a second quantized
world sheet field theory. As we have already emphasized, the interaction
creates and destroys solid lines (boundaries), and therefore second 
quantization becomes indispensable. We follow the ideas introduced in [3]
and define fields that create and destroy boundaries. A crucial difference
is that, in contrast to [3], these fields are labeled by the charges, and not
by the potentials. This enables us to write down a simple local
expression for the free massless action (eq.(9)). 

In section 5, we introduce the boundary changing interaction (eq.(12)).
However, this interaction is not quite correct, since it generates
multiple solid lines which lead to over counting. This problem was overcome
in [7] by introducing world sheet fermions. The interaction is then modified by
fermionic terms (eq.(17)), and Fermi statistics forbids the unallowed
states. The constraint that projects out the unallowed states is
expressed by eq.(21).

There is, however, another problem with the interaction that is not so
easily cured: The prefactor $1/(2 p^{+})$ in eq.(1), which has to be
attached to the interaction, is missing. To try to incorporate this
factor would lead to complicated non-local terms. Instead, we observe
that this factor is needed for Lorentz invariance in the light cone
picture. In the covariant model we will develop, there is no 
 such extra factor. Since we regard the present non-covariant model
only as a laboratory in preparation for the eventual covariant theory,
we will leave its present simple though imperfect form.

So far, we have been dealing with a massless model. In section 6, we
add a simple local  mass term to the action. This is done
by introducing an extra spacelike dimension $s$, and coupling it to
electric dipoles located on the boundaries. The dipole moment of these
dipoles is proportional to the mass. This completes the electrostatic
analogy: The charges on the boundaries generate the momentum
dependent term in the propagator, and the dipoles generate the mass term.
However, the equations of motion for $s$ (eq.(26)), in addition to
the desired mass term, allow for unwanted solutions. We show how these
extra solutions can be projected out by means of an orbifold
projection (eq.(27)).

Starting with section 7, we will develop the covariant world sheet
theory that has been our goal. We start with the first quantized 
picture of the world sheet. The first step is to cast
 the free massless light cone action of eq.(2) into a covariant form.
This was already done in [2], with the result  given by eq.(28). In section 7,
we present a  review of the proof of equivalence of the covariant
and the light cone actions given in  [2]. The proof consists
of showing that the additional field $\lambda$ that was introduced in
the covariant version can be set equal to one. At the same time, the
coordinates $\sigma$ and $\tau$ can be fixed at their light cone values
$p^{+}$ and $x^{+}$, completing the proof of equivalence.
 Invariance of the action under restricted $\sigma$
and  $\tau$ reparametrizations play an important role in reaching this
result. For the second quantized covariant
 version of the model, we only need to
fix $\lambda=1$; the coordinates $\sigma$ and $\tau$ are left
arbitrary. Finally, we also give the covariant version of the mass term.

In section 8, we show how to convert the first quantized covariant theory
(eqs.(52) and (53)) into a second quantized one. This amounts to
almost repeating the derivation given for the light cone version in
sections 4 and 5. Essentially, all one has to do is to replace the
transverse momenta $\cue$ and the charge $\zhee$ by the covariant
vectors $q^{\mu}$ and $z^{\mu}$. This leads to the main result of this
paper: A covariant second quantized theory, with the action (62),
supplemented by constraints on the states (eq.(61). Finally, we argue
 that the constraint resulting from varying the action with respect to
$\lambda$ should be reintroduced in a suitably weakened form. 
Section 9 summarizes our conclusions and possible future directions
of research.

\vskip 9pt
\noindent{\bf 2. The Light Cone World Sheet Picture}

\vskip 9pt

We start with a brief review of the  representation of the
planar graphs of the $\phi^{3}$ theory in the light cone variables
on the world sheet [1].
Although in this paper we are mainly interested  in
covariant Feynman graphs, we do not know how to cover the world sheet
of the interacting theory with these graphs directly
 in a manifestly covariant picture. Instead, we choose an indirect
route:  We first ansatz
a covariant world sheet theory, and then
 by suitably fixing coordinates, show that
it is equivalent to the well established light cone picture. In this picture,
the world sheet is parametrized by the light cone coordinates 
$\tau=x^{+}$ and $\sigma=p^{+}$ as a collection of solid lines (Fig.1),
where the n'th line carries a $D-2$ dimensional transverse momentum
${\bf q}_{n}$. Two adjacent solid lines labeled by n and n+1 correspond
to the light cone propagator
\be
\Delta({\bf p}_{n}) =\frac{\theta(\tau)}{2 p^{+}_{n}}\,\exp\left(-i \tau
\,\frac{{\bf p}_{n}^{2} + m^{2}}{ 2 p^{+}_{n}}\right),
\ee
where ${\bf p}_{n}=\cue_{n} -\cue_{n+1}$ is the transverse momentum flowing
through the propagator, and the width of the strip is
 $p^{+}_{n}$, the $+$ component of the
momentum carried by the propagator.
 A factor of the coupling constant g is inserted
at the beginning an at the end of each line, where the interaction
takes place. Ultimately, one has to integrate over all possible
locations and lengths of the solid lines, as well as over the
momenta they carry. We note that the cubic interaction vertex is the
same as the corresponding open string vertex in the light cone picture.

There is a spurious singularity peculiar to the light cone picture
at $p^{+}=0$. One way to avoid this singularity is to discretize
the coordinate $\sigma$ in steps of length $a$. A useful way of visualizing
the discretized world sheet is pictured in Fig.2.
\begin{figure}[t]
\centerline{\epsfig{file=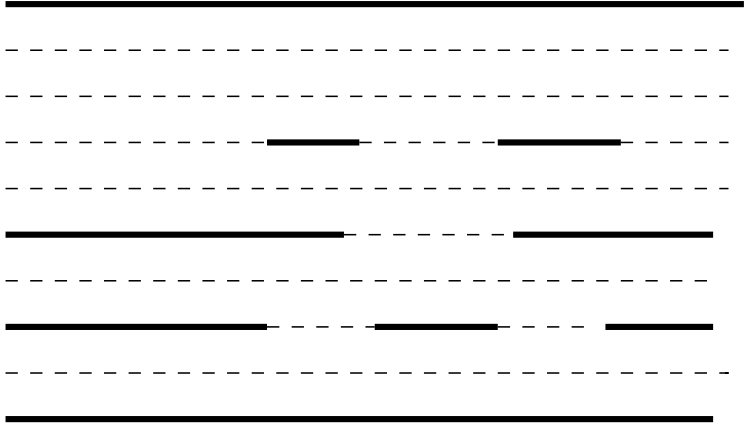, width=10cm}}
\caption{Solid And Dotted Lines}
\end{figure}
 The boundaries of the
propagators are marked by solid lines, and the bulk is filled by dotted lines
spaced at a distance $a$.
 In the covariant picture,
there is of course no  singularity at $p^{+}=0$, and we will mostly
take $\sigma$ to be continuous, although
  we still occasionally use
discretization as a convenient way to describe the world sheet. In this
article, the world sheet field theory will be treated classically, and
so the formal passage from discrete to continuous $\sigma$ will present
no problem. Of course, if higher order corrections to the classical
picture are considered, there will in general be divergences which need
to be renormalized. One has then to introduce some kind of regulator,
possibly the discretized world sheet or something else to
 study the continuum limit more carefully. We hope to address this
important problem in the future.

\vskip 9pt

\noindent{\bf 3. Light Cone World Sheet Field Theory
For The Free Massless Scalar}

\vskip 9pt

In this and the next
 section, we will develop a world sheet field theory which will
reproduce the light cone graphs described earlier, with the exception
of the prefactor $1/(2 p^{+})$. This will serve as a stepping stone
for the Lorentz covariant theory which is the focus of our interest.
We will see that having understood this simplified model, it will be
easy to generalize it to a fully covariant theory. We first briefly
review the world sheet formalism developed in [2]. Consider first in the 
light cone variables the non-interacting massless model,  which is a
collection of infinitely extended propagators. The corresponding world
sheet  action is given by
\be
S_{0}=\intsigt \left(- \frac{1}{2} (\parsig \cue)^{2} +
\sum_{n} \delta(\sigma -\sigma_{n})\,\bfwh_{n}\cdot \parta \cue \right).
\ee
Here $\sigma_{n}$ mark the boundaries of these propagators (solid lines),
and $\bfwh_{n}$ are Lagrange multipliers ensuring that the
momentum $\cue(\sigma_{n}, \tau)= \cue_{n}(\tau)$,
 flowing through the n'th solid line
is conserved in time. Solving the classical equation of motion
\be
\parsig^{2} \cue= \sum_{n} \delta(\sigma -\sigma_{n})\,\parta\bfwh_{n}
\ee
for $\cue(\sigma,\tau)$ in the interval $\sigma_{n}\leq \sigma \leq
\sigma_{n+1}$ in terms of its $\tau$ independent boundary values
$\cue_{n}=\cue(\sigma_{n})$ leads to the result
\be
\cue(\sigta)=\frac{(\sigma -\sigma_{n})\, \cue_{n+1} +(\sigma_{n+1} -\sigma)
\,\cue_{n}}{\sigma_{n+1} -\sigma_{n}}.
\ee
Substituting this in the action gives
\be
S_{0}\rightarrow -\frac{1}{2}(\tau_{f}-\tau_{i})\,\sum_{n}
\frac{\left(\cue_{n+1}-\cue_{n}\right)^{2}}{\sigma_{n+1}-\sigma_{n}}.\ee
Identifying 
$$
\sigma_{n+1}-\sigma_{n}=p_{n}^{+},\,\,\,\cue_{n+1}-\cue_{n}={\bf p}_{n},
$$
and $\tau_{i,f}$ with the initial(final) time, we recover eq.(1).

These equations admit a useful one dimensional electrostatic analogy.
The function $\cue$ can be identified with the electrostatic potential,
and after integrating by parts with respect to $\tau$ in eq.(2),
\be
\zhee_{n}= -\parta\bfwh_{n}
\ee
can be identified
with line charges located at on the boundaries at $\sigma=\sigma_{n}$.
We note that these charges are conserved:
$$
\parta \zhee_{n}=0.
$$

If so desired, both $\cue$ and the action can be expressed in terms
of the charges. For example, from eq.(3)
\be
\cue(\sigta)= \frac{1}{2} \sum_{n} |\sigma- \sigma_{n}|\,\zhee_{n}(\tau).
\ee
At this point, one is free to choose either the the potentials
$\cue(\sigma_{n}, \tau)$ or the charges $\zhee_{n}$ as independent
dynamical variables. Using the above set of equations, it is easy to
 transform 
from one set of variables to the other set.

\vskip 9pt

\noindent{\bf 4. Second Quantization}

\vskip 9pt

The problem with the  formalism described above is that it is  difficult to 
introduce the interaction. We recall that the interaction takes place where a
solid line begins or ends, and the picture developed so far cannot 
accommodate creation or destruction of the solid lines (boundaries). For
this purpose, we need a second quantized theory with fields that
create and destroy the boundaries. Such a second quantized theory was
introduced in [3] and developed further in [4].
 Here, we will make use of the same idea, with however,
an important difference. In [3], the fields were labeled, in addition to
 $\sigma$ and $\tau$, by the momenta (potentials) $\cue$. Here, instead
of $\cue$, we will label the fields by the charge $\zhee$. Accordingly,
we introduce a complex scalar field and its conjugate, which
satisfy the commutation relations
\be
[\phi(\sigta,\zhee), \phi^{\dagger}(\sigtap,\zhee')]=
\delta(\sigma -\sigma')\,\delta(\zhee -\zhee').
\ee
The field $\phi$ destroys a solid line at $\sigma$ carrying a charge
$\zhee$ and $\phi^{\dagger}$ creates such a line. Vacuum is the empty
world sheet (all dotted lines), annihilated by the $\phi$'s.

The motivation for switching from $\cue$ to $\zhee$ in labeling fields
is to  try to simplify the resulting field theory. In reference [3],
the world sheet field theory in the $\cue$ basis contained complicated
non-local terms. We have found it  difficult to convert these terms
into a manifestly covariant form. In contrast, in $\zhee$ basis, a simple
local theory emerges, and this can easily be generalized to a covariant
form. We start with the second quantized version of eq.(2):
\bq
S_{0}&=& \int d\tau \left(i \int d\sigma \int d\zhee\, \phi^{\dagger}
\parta \phi - H_{0}(\tau)\right),\nonumber\\
H_{0}&=& \int d\sigma \left( \frac{1}{2}(\parsig \cue)^{2}
+\cue \cdot \int d\zhee\, \phi^{\dagger} \zhee \phi\right).
\eq
The $\zhee$ integration is over the $D-2$ dimensional transverse space.

To see that this is the right action, we first note that the first term
in the equation for $S_{0}$ implies the canonical commutation relations
between $\phi$ and $\phi^{\dagger}$ (eq.(8)). Next, consider the state
$|s\rangle$ at some fixed $\tau$,
with a collection of solid lines placed at points $\sigma=\sigma_{n}$:
\be
|s\rangle= \prod_{n} \phi^{\dagger}(\sigma_{n},\zhee_{n})|0\rangle.
\ee
It is easy to show that
\be
\cue \cdot \int d\zhee\, \phi^{\dagger} \zhee \phi\, |s\rangle =
\sum_{n} \delta(\sigma -\sigma_{n})\,\cue \cdot \zhee_{n}|s\rangle,
\ee
and remembering that $\zhee_{n}= -\parta\bfwh_{n}$, $H_{0}$,
acting on $|s\rangle$, exactly
reproduces the action of eq.(2).

There is an important restriction on the state $|s\rangle$: Each site
$\sigma_{n}$ is singly occupied by the corresponding 
$\phi^{\dagger}(\sigma_{n},\zhee_{n})$. This is easier to see
if $\sigma$
is discretized; multiple solid
lines at the same site would lead to overcounting. We will call such
a state an allowed state. In a free model, if we require all the states
at some initial time to be allowed, their time development will still
leave them as allowed states. We will see that things are not so simple
in the presence of interaction.

\vskip 9pt

\noindent{\bf 5. The Interaction}

\vskip 9pt

The simplest way to introduce interaction is to set
\be
H_{I}=\int d \sigma \int d \zhee\,g \left(\phi(\sigma, \zhee)+
\phi^{\dagger}(\sigma, \zhee)\right),
\ee
where g is the coupling constant. To show that this reproduces the
the perturbation expansion, we expand in powers of $g$, using
the interaction representation. In this representation, $\phi$ is a free
field that satisfies the equation
$$
\parta \phi=0,
$$
and the commutation relations (8). Acting on the states (10),
  $\phi^{\dagger}$ creates a solid
line and $\phi$ annihilates one and, expanding in powers of $g$ in the
interaction representation generates the perturbation graphs pictured in
Fig.2. 

However, this interaction
would also create unallowed states with multiple solid lines. One way to 
overcome this problem is to modify the interaction by adding fermionic
fields [7]. We introduce two fermionic fields and their conjugates, which
satisfy the anti commutation relations
\be
[\psi_{i}(\sigta), \psi_{i'}^{\dagger}(\sigtap)]_{+}= \delta_{i,i'}\,
\delta(\sigma-\sigma'),
\ee
with $i,i'=1,2$. The structure of the fermionic states is best visualized when
$\sigma$ is discrete: The site of a dotted line is occupied by one
$\psi^{\dagger}_{1}$ and the site of a solid line by one $\psi^{\dagger}_
{2}$. Thus, the state $|0\rangle$ corresponding to the empty world
sheet (all dotted lines) satisfies
\be
 \psi_{2}|0\rangle= 0,\,\,\,\psi_{1}^{\dagger}|0\rangle=0,\,\,\,
\phi|0\rangle=0,
\ee
for all $\phi$'s and $\psi$'s. We now modify eq.(10) to redefine an allowed
state at a fixed $\tau$:
\be
|s_a\rangle= \prod_{n}\left( \phi^{\dagger}(\sigma_{n},\zhee_{n})\,
\psi^{\dagger}_{2}(\sigma_{n})\right)  |0\rangle.
\ee
Multiply occupied lines are now forbidden by  Fermi statistics.

To ensure that allowed states develop into other allowed states, one
has to modify the interaction (eq.(12)). We define the composite operators
\be
\rho^{+}= \psi_{1}^{\dagger}\,\psi_{2},\,\,\,\rho^{-}=\psi_{2}^{\dagger}\,
\psi_{1},\,\,\,\rho=\psi_{2}^{\dagger} \psi_{2},
\ee
and with their help, we rewrite the interaction as 
\be
H_{I}=\int d\sigma \int d \zhee\, g \left(\rho^{+}(\sigma)\,\phi(\sigma,
\zhee)+ \rho^{-}(\sigma)\,\phi^{\dagger}(\sigma,\zhee)\right).
\ee
This new interaction hamiltonian, as well as the total hamiltonian
\be
H=H_{0}+ H_{I}
\ee
 maps allowed states into other
allowed states. One way to see this is to notice that the
operator
\be
K(\sigta)= \int d \zhee\, \phi^{\dagger}(\sigta, \zhee)\,
\phi(\sigta, \zhee) -\rho(\sigta)
\ee
commutes with H,
\be
[K,H]=0
\ee
and annihilates the allowed states:
\be
K |s_{a}\rangle=0.
\ee
Here, we have made use of
\be
\rho(\sigma)|s_{a}\rangle =\left( \sum_{n} \delta(\sigma-\sigma_{n})
\right)|s_{a}\rangle.
\ee
Eq.(21), imposed at a fixed $\tau$, can serve as the definition of the
allowed states at that $\tau$. Since $K$ commutes with $H$, this condition 
then holds for all $\tau$.

In [4], we imposed a condition analogous to (21), with $\zhee$ replaced by 
$\cue$. Actually, we had a stronger version of it
$$
K=0,
$$
which was implemented by multiplying with a Lagrange multiplier and adding
it to the action. This has certain computational advantages, but here
we adopt the weaker condition (21), and impose it as a boundary condition
on the initial states.

The action including the fermions and the interaction is given by
\be
S=\int d\tau\left(i \int d\sigma\left(\psi^{\dagger}\parta \psi+
\int d\zhee\,\phi^{\dagger} \parta \phi \right) -H(\tau)\right),
\ee
with $H$ given by eq.(18).

This action has an important advantage over the previous
 version given in
references [3,4]: It is local in both the coordinates $\sigma$ and $\tau$.
We recall that the non-locality in [3] had two sources: The propagator
must be attached to two adjacent solid lines; non-adjacent solid lines
do not generate a propagator. This is because
 in the $\cue$ (potential)
basis, the propagator is generated by a
 potential that satisfies boundary conditions  on the
adjacent solid lines.
 In [3], this condition was implemented by attaching
a factor non-local in $\sigma$ which vanishes for non-adjacent solid lines.
This problem is  avoided in the charge (\zhee) basis: All 
charges, adjacent or non-adjacent, interact with the same one
dimensional coulomb potential
$$
\frac{1}{2}\,|\sigma -\sigma'|.
$$

There is, however, one remaining problem: 
The prefactor $1/(2 p^{+})$ in
eq.(1) is missing. In [3, 4], this factor was attached
 to the interaction, and if we tried the same
approach here, this would again have to
 introduce complicated non-local terms.
Instead, as we will discuss in more detail later, $1/(2 p^{+})$ is a
measure factor needed in switching from the covariant propagator to
the light cone propagator.
 In the manifestly Lorentz invariant
version of the model, which will be the focus of our interest, there is
no such extra factor, and this complication does not arise. Since we are using
the above non-covariant model as an introduction to
 the covariant theory, we will leave
it in its present simple though imperfect form.

\vskip 9pt

\noindent{\bf 6.  The Mass Term}

\vskip 9pt

The model developed so far is massless. To introduce a finite mass term,
we will add an extra space like dimension to the model. We could try
compactifying the extra dimension, but this would generate a whole
tower of Kaluza-Klein states. In order to generate a single massive
particle, we again make use of the electrostatic analogy. Let the
potential corresponding to the extra dimension be $s(\sigta)$, and let the
solid lines be located at $\sigma_{n}$. In
 eq.(4), we make the replacements
$$
\cue\rightarrow s,\,\,\,\cue_{n+1}\rightarrow m/2,\,\,\,
\cue_{n} \rightarrow - m/2,
$$
and in the interval $\sigma_{n}\leq \sigma \leq \sigma_{n+1}$, we have
\be
s(\sigma)=\frac{m}{2}\,\frac{2 \sigma -\sigma_{n}-\sigma_{n+1}}
{\sigma_{n+1}-\sigma_{n}},
\ee
where $m$ is the mass. The resulting saw-tooth configuration is sketched in
Fig.3.
\begin{figure}[t]
\centerline{\epsfig{file=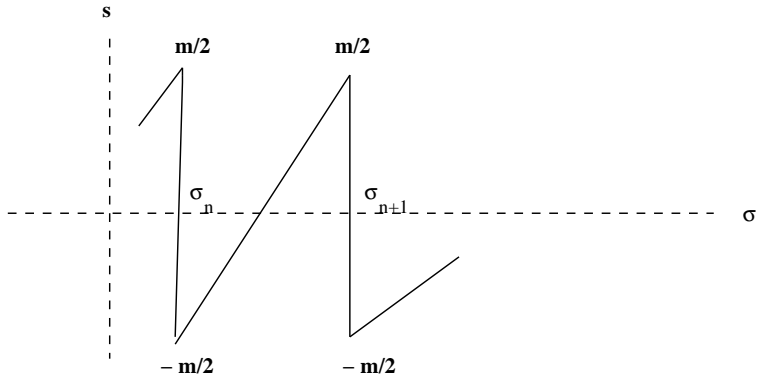, width=10cm}}
\caption{The Saw-tooth Configuration}
\end{figure}
 It is easy to see that the corresponding action is again eq.(5),
with, however, the replacement
$$
(\cue_{n+1}- \cue_{n})^{2}\rightarrow m^{2}.
$$
This is precisely the mass term in the propagator (1).

To construct the corresponding action, let us recall
that in the case of the massless model, we placed charges on the solid
lines, and the resulting potential reproduced the light cone propagator. The 
potential $\cue$ was continuous across a solid line, and the electric field
jumped by minus the charge $\zhee$. In contrast, $s$ is discontinuous at
$\sigma=\sigma_{n}$, with a jump given by $m$. This corresponds, instead of
charge, to a dipole of strength $m$. A simple action which takes the dipole
sources into account is
\be
S_{m}=\intsigt \left( -\frac{1}{2}( \parsig s)^{2} - m\,s\,\parsig \rho
\right),
\ee
leading to the equation of motion
\be
\parsig^{2} s= m\,\parsig \rho
= m\, \sum_{n} \delta'(\sigma -\sigma_{n}),
\ee
which follows from eq.(22). This equation of motion admits the $s$ given by
(4) as a solution, and as we have already argued,
 this solution, substituted into the action
reproduces the mass term in the propagator (1).

There is, however, still a remaining problem; although $s$ given by
(24) is a solution to (26), it is not the unique solution. 
The equation of motion for $s$
 only requires that crossing a solid line in the
positive direction, $s$ jumps by a total amount $m$, say from $m_{1}$ to
$m_{1}+m$. It does not require it to jump from  $- m/2$ to $+ m/2$;
 $m_{1}= - m/2$ has to be imposed as an additional boundary
condition. We have found that the simplest way to do this is by an
orbifold projection. The action (25) is invariant under the $Z_{2}$
symmetry
\be
s(\sigma)\rightarrow - s(-\sigma),\,\,\, \rho(\sigma)\rightarrow
\rho(- \sigma),
\ee
assuming that the range of integration is symmetric under the reflection
of $\sigma$.
We now require  the solutions to equations of motion (26) to be also
invariant under this symmetry. From Fig.3, the saw-tooth configuration
is clearly invariant under (27): Running $\sigma$ backwards
changes the sign of $s$. It is not difficult to show that the converse
is also true: Symmetry under (27) fixes the solution to the
equations of motion  to be
the saw-tooth; all other configurations are projected out.

\vskip 9pt

\noindent{\bf 7. Covariant World Sheet Field Theory For The
Free Scalar}

\vskip 9pt

In this section, we will introduce a covariant generalization of the
first quantized light cone world sheet action (eq.(2)), and show that it is
correct by transforming it into its light
cone version. This was first done in section 3 of [2], and so
this section is mostly a review. We will, however, have
 some additions and amplifications; for example, the mass term (eq.(53)) is
new.
 We start with the
the covariant action given in [2] :
\be 
S_{0}=\intsigt \left(\frac{1}{2} \lambda\,\parsig q^{\mu} \parsig q_{\mu}
+ \sum_{n}\delta(\sigma -\sigma_{n})\, y_{n}^{\mu} \parta q_{\mu}
\right).
\ee
Comparing this with the light cone action (2), we notice the following:\\
a) As opposed to the transverse $\cue$, $q^{\mu}$ and $y_{n}^{\mu}$
are  Minkowski vectors in $D$ dimensions. As before, $y_{n}$'s are
the Lagrange multipliers for the conservation equations
\be
\parta q^{\mu}(\sigma_{n},\tau)=0.
\ee
b) The new field $\lambda(\sigta)$ looks like a Lagrange multiplier.
However, we shall see that it is closely related to the Schwinger
proper time. It is therefore restricted to be positive semi-definite.\\
c) In the light cone action, $\sigma$ and $\tau$ are identified with
$p^{+}$ and $x^{+}$ respectively. Here, $\sigma$ and $\tau$ are arbitrary
parameters.

We will now proceed as at the beginning of section 3 by
 solving the equations of motion for $q^{\mu}$ in terms of $\tau$
independent vectors $q_{n}^{\mu}=q^{\mu}(\sigma_{n})$. If $\sigma$ lies in
 the interval $\sigma_{n}\leq \sigma \leq \sigma_{n+1}$, then
\be
 q^{\mu}(\sigta)= q^{\mu}_{n} + v^{\mu}_{n}(\tau)\,\int^{\sigma}_
{\sigma_{n}} d\sigma'\,\lambda^{-1}(\sigtap).
\ee
Setting $\sigma=\sigma_{n+1}$ gives
\be
 v^{\mu}_{n}(\tau)=\frac{q^{\mu}_{n+1} - q^{\mu}_{n}}{\kappa_{n}(\tau)},
\ee
where we have defined
\be
\kappa_{n}(\tau)=\int^{\sigma_{n+1}}_
{\sigma_{n}} d\sigma'\,\lambda^{-1}(\sigtap).
\ee
Using these results. the  action (28) becomes,
\bq
S_{0}&=&\int d\tau \sum_{n} \int^{\sigma_{n+1}}_{\sigma_{n}}
 d\sigma\left(
\frac{1}{2} \lambda\,\parsig q^{\mu} \parsig q_{\mu}\right)\nonumber\\
&=& \int d\tau\,\sum_{n}
 \frac{\left(q^{\mu}_{n+1}- q^{\mu}_{n}\right)^{2}}
{2 \kappa_{n}(\tau)}.
\eq

It is now easy to generalize to the massive case. Eq.(25) is replaced by
\be
S_{m}=\intsigt \left( -\frac{\lambda}{2}( \parsig s)^{2}
 - m\,s\,\sum_{n} \delta'(\sigma -\sigma_{n})\right),
\ee
and with the addition of this term, the action becomes
\be
S_{0}=\int d\tau\,\sum_{n} \frac{\left(q^{\mu}_{n+1}- q^{\mu}_{n}\right)^{2}
- m^{2}}
{2 \kappa_{n}(\tau)}=\int d\tau\,\sum_{n}\frac{(p_{n}^{\mu})^{2}- m^{2}}
{2 \kappa_{n}(\tau)}.
\ee

We note that the final result for the action depends only on
$\kappa_{n}(\tau)$ and not on the full $\lambda(\sigta)$; the $\sigma$
dependence has been integrated over. This a kind of gauge invariance:
We can shift $\lambda^{-1}$ by an arbitrary function of $\sigma$ and
$\tau$, so long as we keep $\kappa_{n}$ fixed. By suitable gauge fixing, 
it is then possible to make $\lambda$ independent of $\sigma$ for each
propagator separately.

Let us now focus on the contribution of a
 single interval in this action, say $\sigma_{n}\leq \sigma \leq
 \sigma_{n+1}$; this term is invariant under the $\tau$ reparametrization
\be
q^{\mu}(\sigta)\rightarrow q^{\mu}(\sigma, f_{n}(\tau)),\,\,\,
\lambda(\sigta)\rightarrow f'_{n}(\tau)\,\lambda(\sigma, f_{n}(\tau)),
\,\,\,\kappa_{n}(\tau)\rightarrow \frac{\kappa_{n}(f_{n}(\tau))}
{f'_{n}(\tau)}.
\ee
recalling that $\lambda$ and $\kappa$ are restricted to be positive,
we also require
$$
f'(\tau)\geq 0.
$$
We will now assume that subject to these restrictions, it is possible
to choose the $f_{n}$'s so as to map the $\kappa_{n}$'s into unity:
\be
\kappa_{n}(\tau)\rightarrow \frac{\kappa_{n}(f_{n}(\tau))}
{f'_{n}(\tau)}=1.
\ee
If now the range of $\tau$ in the n'th term is from 0 to $T_{n}$, then,
integrating over $T_{n}$,
\be
\int_{0}^{\infty} d T_{n}\,\exp\left( i\int_{0}^{T_{n}}\left((p_{n}^{2}
-m^{2}\right)\right)=\frac{i}{p_{n}^{2} -m^{2} -i\epsilon},
\ee
we get the covariant Feynman propagator. Here $T$ can be identified
with the Schwinger proper time, after a Euclidean rotation.

The above derivation of the covariant propagator is somewhat heuristic; a more
careful discussion of the $\tau$ reparametrization fixing is needed. 
Rather then trying to clean up this derivation, we will instead review
a better founded derivation of the light cone propagator given in [2].
As pointed out by Thorn [5], once the light cone propagator is established,
one can switch to the Schwinger representation by means of the
transformation
\bq
&&\int \frac{d p^{-}}{2 \pi}\, e^{- i x^{+}\,p^{-}} \int d T \exp\left(
- i T\left({\bf p}^{2} -2 p^{+}\,p^{-} + m^{2} -i\epsilon \right)\right)
\nonumber\\
&=&\int d T\,\delta\left(x^{+} -2 T\,p^{+}\right)\,e^{- T ({\bf p}^{2}
+ m^{2})},
\eq
where $x^{+}$ is identified with the time $\tau$. The prefactor 
$1/2 p^{+}$  comes from changing the variable of integration
from $T$ to $x^{+}=\tau$.

In the covariant approach given above, the $y$'s did not appear
explicitly in the action; instead, the action was expressed
in terms of the boundary values of the $q$'s which are $\tau$ independent.
 We will now derive a new representation for the action, which
is better suited to the light cone picture. In this representation, a mixed
set of variables are used. First,
transverse ${\bf y}_{n}$'s and $y_{n}^{-}$'s are integrated over, resulting
in the conservation equations
\be
\parta \cue_{n}(\tau)=0,\,\,\,\parta q_{n}^{+}=0.
\ee
The action can now be written in terms of the above set of 
$\tau$ independent variables, plus $y_{n}^{+}(\tau)$ and $q^{-}(\sigta)$:
\be
S_{0}=\intsigt\left(-\frac{1}{2} \lambda (\parsig \cue)^{2} +\lambda\,
\parsig q^{+}\,\parsig q^{-} -\sum_{n}\delta(\sigma -\sigma_{n})\,
q^{-}_{n}\,\parta y_{n}^{+}\right).
\ee
We notice that this action is invariant under the following
 reparametrization:
\be
q^{\mu}(\sigta)\rightarrow q^{\mu}(h(\sigta), \tau),\,\,\,
\lambda(\sigta)\rightarrow \left(\parsig h(\sigta)\right)^{-1}\,
\lambda(h(\sigta), \tau),
\ee
if $h$ satisfies the conditions
$$
h(\sigma_{n},\tau)=\sigma_{n},
$$
so that the boundaries are kept straight and the last term in eq.(41)
stays invariant.

We now make use of this invariance to cast (41) into the light cone action
(2). First, transforming by means of a $\tau$ independent $h$, we can set
\be
\sigma_{n+1}- \sigma_{n}= q_{n+1}^{+} -q_{n}^{+}= p_{n}^{+}
\ee
at some fixed $\tau=\tau_{0}$. Since both sides of this equation are
$\tau$ independent (see eq.(40)), it is then valid for all $\tau$. By fixing 
one $\sigma$ to be zero
 at one boundary, we might as well set
\be
 \sigma_{n+1}=
q^{+}_{n+1}=p_{n}^{+}.
\ee

Next, transforming by a $\tau$ dependent $h$ that preserves the boundaries,
we can set
$$
\sigma=q^{+}(\sigta)\rightarrow \parsig q^{+}=1
$$
in the bulk also, and not just on the boundaries. Substituting this in (41),
the equation of motion for $q^{-}$ in the bulk leads to
\be
\parsig \lambda=0,
\ee
so $\lambda$ depends only on $\tau$, a result which we have already derived
 in a different way following eq.(35). This result only holds in the bulk;
using the equations of motion with respect to $q^{-}$ gives the jump of
$\lambda$ across the boundaries:
\be
\lambda(\sigma_{n}+\epsilon ,\tau) - \lambda(\sigma_{n}-\epsilon, \tau)
=\parta y^{+}_{n}.
\ee

 We now appeal the $\tau$ reparametrization invariance (eq.(36)). By
suitably choosing different $f_{n}$'s, we can completely eliminate
the discontinuities in $\lambda$, and in view of (45), $\lambda$ can be
set to a constant over the whole world sheet. Without loss of generality,
this constant can be taken to be one:
\be
\lambda=1.
\ee
From eq.(46), we also have,
\be
\parta y_{n}^{+}=0.
\ee
At this point, we need to relate the y's to the position coordinate
$x^{\mu}(\sigta)$.
It was shown in [2] that  $x^{\mu}(\sigta)$ is $\sigma$ independent
in the bulk, and its jump over the n'th boundary is $y_{n}^{\mu}$.
This is summarized by the equation
\be
\parsig x^{\mu}(\sigta)=\sum_{n} \delta(\sigma -\sigma_{n})\,
y_{n}^{\mu}(\tau),
\ee
which can be taken as the definition of $x^{\mu}$.

Now, it  follows from (48) and (49)
that $\parta x^{+}$ has no jumps on the boundaries, and therefore it
 is completely $\sigma$ independent.  Using an overall $\tau$
reparametrization still at our disposal, we can therefore set
\be
\parta x^{+}=1 \rightarrow x^{+}=\tau.
\ee

To summarize, we have shown that, by suitable $\sigma$ and $\tau$
 reparametrizations, one can impose the conditions
\be
\lambda=1,\,\,\,\sigma=p^{+},\,\,\,\tau=x^{+},
\ee
and these are all that is needed to map the covariant theory into the
light cone model of the previous sections. We have not so far discussed the 
mass term, but it is clear that $\lambda=1$
is sufficient to show the equivalence of the light cone and the
covariant mass term . Of course, we still have to orbifold
by the $Z_{2}$ symmetry of (27).

Although the constraint $\lambda=1$ came at the end of a lengthy process of
fixing the coordinates, it could have been derived more simply right at 
the beginning by appealing only to $\tau$ reparametrization. We recall that
the action does not depend on the full $\lambda$ but only the 
$\kappa_{n}(\tau)$'s (eq.(35)). These can be set equal to unity 
by the $\tau$ reparametrizations (eq.(37)). It is important to note that,
by leaving the $\sigma$ coordinate and the overall
$\tau$ reparametrization arbitrary,
this can be done without imposing the conditions
$$
\sigma=p^{+},\,\,\,\tau=x^{+},
$$
 By suitable $\tau$
reparametrization, we can therefore replace eq.(28) by
\be 
S_{0}=\intsigt \left(\frac{1}{2} \parsig q^{\mu}\, \parsig q_{\mu}
+ \sum_{n}\delta(\sigma -\sigma_{n})\, y_{n}^{\mu}\, \parta q_{\mu}
\right),
\ee
and similarly, the mass term becomes
\be
S_{m}=\intsigt \left( -\frac{1}{2}( \parsig s)^{2}
 - m\,s\,\sum_{n} \delta'(\sigma -\sigma_{n})\right).
\ee

These expressions for $S_{0}$ and $S_{m}$ will be the starting point
for the covariant quantization in the next section. We note that
they are manifestly covariant, and they can be obtained from
the analogous light cone expressions by the simple
substitution
\be
\cue\rightarrow q^{\mu},\,\,\,{\bf y}\rightarrow y^{\mu}.
\ee
For example, in analogy with eq.(6), we  define the covariant charge
by
\be
z^{\mu}= -\parta y^{\mu},
\ee
and note that this charge is also conserved :
$$
\parta z^{\mu}=0.
$$
Eq.(7) also has its obvious covariant analogue. 

The advantage of starting second quantization with eqs.(52) and
(53)  should now
be clear: We can simply repeat the developments of sections 4, 5 and
6, with the obvious replacement of $\cue$ and $\zhee$ by $q^{\mu}$
and $z^{\mu}$. There is, however, one remaining problem that needs to be
clarified. By setting $\lambda=1$, the equations of motion with respect
to $\lambda$ are lost. We already pointed out that these equations are
too stringent; but we cannot just ignore them. In the next section,
we will argue that
they have to be replaced by weaker set of constraints.

Finally,  a few comments on the $1/(2 p^{+})$
factors. These are guaranteed to appear in the light cone picture
by the requirement of Lorentz invariance. In fixing the $\sigma$
coordinate by eq.(44), there is a possible measure factor (Jacobian)
in the functional integral which we have neglected. We have not
directly computed this factor; but indirectly, Lorentz invariance
tells us that it should be proportional to $1/p^{+}$.
 In fact, we have already seen (eq.(39)) that transforming the covariant
propagator into the light cone variables accounts for this factor.

\vskip 9pt

\noindent{\bf 8. Second Quantized Covariant World Sheet Field
Theory}

\vskip 9pt

In the last section, we have seen that the action
$S_{0}+ S_{m}$, generates a 
covariant first quantized free world sheet theory, which, upon
fixing the coordinates, reduces to its light cone version. As we
have pointed out earlier, it is difficult to introduce interaction in
the first quantized theory, and so, following section 4, we second
quantize the model. 
 Actually, our job is easy; having worked out the
non-covariant version in detail, all we have to do is guess the obvious
covariant generalization. We start  with eqs.(52) and (53) for the free
first quantized action, and observe that, their second quantized
version is the analogue of eq.(9):
\bq
S_{0}&=& \int d\tau \left(i \int d\sigma \int dz\, \phi^{\dagger}
\parta \phi - H_{0}(\tau)\right),\nonumber\\
H_{0}&=& \int d\sigma \left(-\frac{1}{2}(\parsig q^{\mu})^{2}
-q^{\mu} \cdot \int dz\,\phi^{\dagger} z_{\mu} \phi\right),
\eq
where the z integration over the full D dimensional space.
 $S_{m}$  is the same as in eq.(25).

The field $\phi$ now depends on $z^{\mu}$ instead of $\zhee$, and satisfies
the commutation relations
\be
[\phi(\sigta, z), \phi^{\dagger}(\sigtap, z')]=
\delta(\sigma -\sigma')\,\delta^{D}(z - z').
\ee
The definitions of the fermions and the $\rho$'s given in section 5 are
unchanged. The definition of the allowed states (eq.(15)) goes over to
\be
|s_a\rangle= \prod_{n}\left( \phi^{\dagger}(\sigma_{n},z_{n})\,
\psi^{\dagger}_{2}(\sigma_{n})\right)  |0\rangle,
\ee
and the interaction Hamiltonian is the same as in eq.(17):
\be
H_{I}=\int d\sigma \int dz\, g \left(\rho^{+}(\sigma)\,\phi(\sigma,
z)+ \rho^{-}(\sigma)\,\phi^{\dagger}(\sigma,z)\right).
\ee
The operator K (eq.(19)) is now defined by
\be
K(\sigta)= \int d z\, \phi^{\dagger}(\sigta, z)\,
\phi(\sigta, z) -\rho(\sigta).
\ee
As before, it commutes with the Hamiltonian and annihilates the
allowed states:
\be
K|s_{a}\rangle =0.
\ee
Conversely, this can be taken as the definition of allowed states.

Finally, the covariant action is 
\be
S=\int d\tau\left(i \int d\sigma\left(\psi^{\dagger}\parta \psi+
\int dz\,\phi^{\dagger} \parta \phi \right) -H(\tau)\right),
\ee
where,
$$
H= H_{0}+ H_{m}+ H_{I}.
$$
 $H_{0}$ and $H_{I}$ are given above, and
$$
H_{m}=\int d\sigma \left( \frac{1}{2}( \parsig s)^{2} + m\,s\,\parsig \rho
\right).
$$

It is important to understand that the action is not the whole story:
The constraints on the states are also crucial. They are restricted
to be the allowed states (58), so they have to satisfy the constraint (61).
 In addition, the equations of motion for $s$ in the mass term should be
solved subject to the orbifold projection (27). The above covariant
action, combined with the also covariant restrictions on the states,
is the main result of this paper. In addition to being manifestly
Lorentz invariant, it has the advantage of being local in the
world sheet coordinates.

How do we know that this is the correct covariant field theory?
To show this, we essentially follow the same steps as in sections 5
and 6. Using the
interaction representation, we expand the hamiltonian
 in powers of the coupling constant $g$. Next, as in section 4,
 we consider the action of the free Hamiltonian on a typical allowed
state. It is easy to show that the analogue of eq.(11),
\be
q^{\mu} \cdot \int dz\,\phi^{\dagger} z_{\mu} \phi\, |s\rangle= 
 \sum_{n} \delta(\sigma -\sigma_{n})\,q_{\mu}\, z^{\mu}_{n}|s\rangle,
\ee
follows. Identifying $z^{\mu}= -\parta y^{\mu}$, this reproduces the 
covariant first quantized free action of eq.(52). The mass term is also
easily seen to reproduce eq.(53). The interaction Hamiltonian, acting on 
states, either creates or annihilates a solid line (boundary) (Fig.2).
 The fermionic
factor ensures that allowed states are mapped into other allowed
states. All of this is in complete parallel with section 5. Once the covariant
first quantized
 world sheet picture is established, we appeal to section 7 to
show that it is equivalent to the light cone picture, which was the
starting point. This completes the proof of equivalence between the covariant
and the light cone theories. Once more, we stress that the restriction
to the allowed states was crucial for the success of this proof of
equivalence.

Finally, we would like to return to the constraint equation
\be
L(\sigta)=\left(\parsig q^{\mu}\right)^{2} - (\parsig s)^{2}=0,
\ee
obtained by varying with respect $\lambda$. It was pointed out in section 7
 that,
as an operator equation, this is too stringent: Going back to eqs.(4) and
(24), in the interval $\sigma_{n}\leq \sigma \leq \sigma_{n}$,
$$
\parsig q^{\mu}= \frac{p_{n}^{\mu}}{\sigma_{n+1} -\sigma_{n}},
\,\,\,\parsig s=\frac{m}{\sigma_{n+1} -\sigma_{n}},
$$
and, therefore,
\be
\left(\parsig q^{\mu}\right)^{2} - (\parsig s)^{2}\rightarrow
\frac{\left(p_{n}^{\mu}\right)^{2} - m^{2}}{(\sigma_{n+1} -\sigma_{n})^{2}}.
\ee
 This leads, for the n'th free propagator, to the mass shell condition
\be
p^{2}_{n}- m^{2}=0,
\ee
instead of the Feynman propagator. In any case, since $\lambda$
is intrinsically positive, we cannot use it as a Lagrange multiplier.
We therefore propose to replace (64) by the following weaker condition:
Instead of vanishing as an operator, L should annihilate the allowed states:
\be
L|s_{a}\rangle =0,
\ee
This only puts the free particle states on the mass shell; it
is therefore quite acceptable. However, we have so far checked
the above constraint only in the context of a perturbation expansion.

 It is quite
possible that in a more general non-perturbative treatment, this has
to be replaced by the even weaker condition: Only the positive
frequency components of $L$ may have to be required to
 annihilate the allowed states. Using mean field approximation,
it was shown in [4,8] that a world sheet densely covered with graphs
leads to string formation. This was done starting with the light
cone picture, so the resulting string was a transverse light cone
string. In contrast, if in some approximation, the covariant approach also 
leads to a string, that would be a covariant string, with its usual
negative metric (ghost) modes. The above constraint may then be needed
to eliminate these unphysical modes.

\vskip 9pt

\noindent{\bf 9. Conclusions And Future Directions}

\vskip 9pt

In this article, we have developed a covariant world sheet description
of the planar $\phi^{3}$ theory. Covariance is of course a desirable
feature, and this new approach, in some ways, should be an improvement
over the well developed light cone approach. It remains to be seen
whether an approximation scheme, such as the mean field method, can be
successfully adapted to this new  model. Another important problem for
future research is to try to generalize from $\phi^{3}$ to more
realistic theories, such as gauge theories \footnote{For some initial steps
 taken towards more realistic theories, see [9, 10].}.
 It would very nice
indeed to be able to formulate gauge theories in a manifestly both Lorentz
and gauge invariant form.

\vskip 9pt

\noindent{\bf Acknowledgment}

\vskip 9pt

This work was supported by the Director, Office of Science,
Office of Basic Energy Sciences, of the U.S. Department of Energy under
Contract No. DE-AC02-05CH11231.

\newpage

{\bf References}
\begin{enumerate}
\item G.'t Hooft, Nucl.Phys. {\bf B 72} (1974) 461.
\item K.Bardakci and C.B.Thorn, Nucl.Phys. {\bf B 626} (2002) 287,
hep-th/0110301.
\item K.Bardakci, JHEP {\bf 0810} (2008) 056, arXiv:0808.2959.
\item K.Bardakci, JHEP {\bf 0903} (2009) 088, arXiv:0901.0949.
\item C.B.Thorn, Nucl.Phys. {\bf B 699} (2004) 427, hep-th/0405018.
\item D.Chakrabarti, J.Qiu and C.B.Thorn, Phys.Rev. {\bf D 74}
(2006) 045018, hep-th/0602026.
\item K.Bardakci, Nucl.Phys. {\bf B 677} (2004) 354, hep-th/0308197.
\item K.Bardakci, JHEP {\bf 1110} (2011) 071, arXiv:1107.5324.
\item C.B.Thorn, Nucl.Phys. {\bf B 637} (2002) 272, hep-th/0203167,
G.Gudmundsson, C.B.Thorn and T.A.Tran, Nucl.Phys. {\bf B 649} (2003) 3,
hep-th/0209102.
\item C.B.Thorn and T.A.Tran, Nucl.Phys. {\bf B 667} (2004) 289,
hep-th/0307203.
\end{enumerate}
\end{document}